ate
\documentclass[%
 reprint,
 amsmath,amssymb,
 aps,
]{revtex4-1}
\usepackage[ps2pdf,colorlinks=true]{hyperref}
\usepackage{graphicx}
\usepackage{dcolumn}
\usepackage{bm}

\begin{document}

\preprint{APS/123-QED}

\title{Observation of $\gamma$-band based on two-quasiparticle configuration in $^{70}$Ge}

\author{ M. Kumar Raju$^{1,2}$}
\email{kumar8284@gmail.com}
\author{P. V. Madhusudhana Rao$^{1}$}
\author{ S. Muralithar$^3$}
\author{ R. P. Singh$^3$}
\author{ G. H. Bhat$^{4}$}
\author{ J. A. Sheikh$^{4}$}
\author{S. K. Tandel$^{5}$}
\author{P. Sugathan$^{1, 3}$}
\author{T. Seshi Reddy$^{1}$}
\author{B. V. Thirumala Rao$^{1}$}
\author{R. K. Bhowmik$^{3}$}
\affiliation{$^1$Nuclear Physics Department, Andhra University, Visakhapatnam - 530003, India}
\affiliation{$^2$Department of Physics, University of the Western Cape, P/B X17, Bellville ZA-7535, South Africa}
\affiliation{$^3$Inter University Accelerator Centre, Aruna Asaf Ali Marg, New Delhi - 110067, India}
\affiliation{$^4$Department of Physics, University of Kashmir, Srinagar 190 006, India}
\affiliation{$^{5}$ UM-DAE Centre for Excellence in Basic Sciences, Mumbai-400098, India}

\date{\today}

\begin{abstract}
The structure of $^{70}$Ge has been studied through in-beam gamma ray spectroscopy.  A new band structure is identified that leads to forking of the ground-state band into two excited bands.  Band structures have been investigated using the microscopic triaxial projected  shell model approach.  The observed forking is demonstrated to result from almost simultaneous band crossing of the two-neutron aligned  and the $\gamma$-band built on this two-quasiparticle configuration with the ground-state band.
\end{abstract}


\maketitle
\section{Introduction}
The atomic nucleus is a fascinating quantum-many body system which shows a rich variety of shapes and structures \cite{BM75}.  Major advances in experimental techniques have facilitated these studies of atomic nuclei at extremes of isospin, angular-momentum and excitation energy. These investigations have revealed new structures and phenomena, hitherto, unknown  in nuclear physics.  In nuclear high-spin spectroscopy, band-structures have been observed up to high angular momentum in some of the nuclei and
investigations of these high-spin states probe the predicted modifications of the shell structure and pairing properties with increasing rotational frequency.  In particular, nuclei in the mass range 60 $\leq$ A $\leq$ 70 display a wide range of phenomena, for instance, co-existence of oblate and prolate shapes, shape changes and dramatic variations in band crossing properties have been observed with particle number. 

In most deformed nuclei, the ground-state band is crossed by a two-quasiparticle aligned structure resulting in the well established phenomenon of backbending \cite{EG73}.  The yrast band after band crossing  consists of two-quasiparticle aligned state and the ground-state configuration becomes the excited band.  In several nuclei this band referred to as the yrare band, is observed up to high-spins.  Further, in  some of the nuclei, the forking of ground-state band into two two-quasiparticle structures have also been observed.  For example, in even-even Xe-Ba-Ce nuclei with N=66-76, the ground-state band forks into two distinct band structures based on h$_{11/2}$ two-quasiparticle configurations \cite{wyss}.  Most of these observed bands after forking have been interpreted as two-neutron and two-proton quasiparticle structures which align almost simultaneously.  The forking in these axially-symmetric nuclei has been explained \cite{forking} as resulting from repulsive nature of neutron-proton interaction in high-j intruder orbital $h_{11/2}$ for the particle-hole configuration. In the present work, we report a forking of ground state band in  $^{70}$Ge. This is shown to arise from a $\gamma-$band built on a two-quasiparticle configuration. 

Well developed $\gamma-$bands are known to exist in many transitional nuclei close to the ground-state which  have been investigated using various phenomenological models \cite{VG81,JL88}.  In the framework of microscopic triaxial projected shell model (TPSM) approach \cite{JS99},  these $\gamma-$ bands result from  projection of the K=2 state of the triaxial self-conjugate vacuum configuration.  This state is a superposition of K=0, 2, 4,.... configuration with  K=0 projected state corresponding  to ground-state band. The projections from K=2 and 4 correspond to $\gamma$- and $\gamma\gamma$- bands, respectively \cite{JS14,JS09}.  It has been demonstrated in several studies  that TPSM approach provides an excellent description of the observed $\gamma$-bands in several mass regions \cite{JS14,JS09}.  It is also obvious from this description that not only the ground-state band, but also the quasiparticle-particle excited configurations should have the associated $\gamma$-bands built on them.  The existence of $\gamma$-band on each intrinsic state has been predicted by Bohr and Mottelson quite sometime back \cite{BM75}. 

Low-spin states of $^{70}$Ge were previously investigated through the $(p,p^\prime$), $(n, n^\prime \gamma$), (p,t) and ($^{3}He, d)$  reactions \cite{previous,previous1,pre2,previous3}.  These studies reported the level structure of $^{70}$Ge up to 5.1 MeV excitation energy.  Later, high spin states were studied by two groups \cite{budda, sugawara}, identified the ground state positive-parity band up to J$^{\pi}$ = (12$^+$) state. In this article, we have presented the  experimental observation of  $\gamma-$band structure built on  two-quasiparticle configuration in $^{70}$Ge.   Experimental details and relevant results are described in Sec.~\ref{exp.details}. Deduced band structures are discussed in Sec.~\ref{discussion} using cranked Hartree Fock Bogoliubov model (CHFBM) and triaxial projected shell model (TPSM) approaches. A brief summary is presented in Sec.~\ref{summary}.

\begin{figure}
\begin{center}
\includegraphics[scale = 0.62, angle=0]{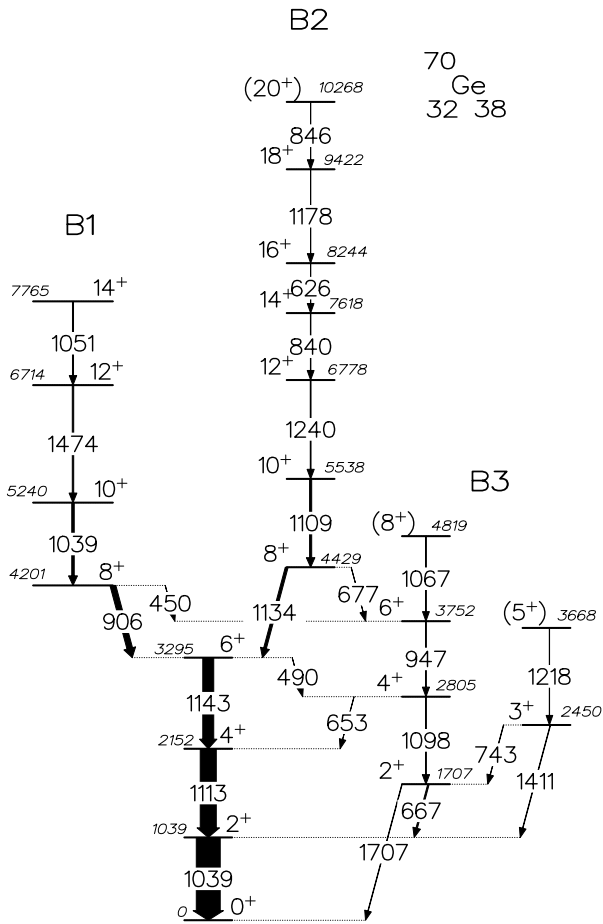}
\caption{Partial level scheme of $^{70}$Ge obtained from the present work. Bands are labeled as B1, B2 and B3 for reference in the text.}\label{fig:level-scheme}
\end{center}
\end{figure} 

\begin{figure}
\includegraphics[width=8.0cm, height=5.6cm, trim = 1cm 0cm 1cm 0cm ]{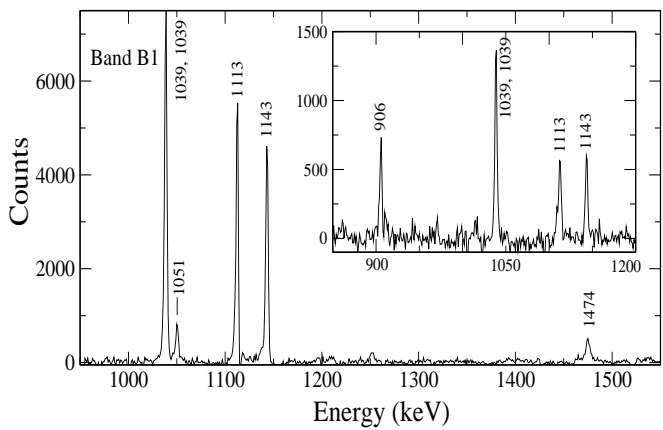}
\caption{\label{band1} A $\gamma$-$\gamma$ coincidence spectrum with a gate on 906-keV $\gamma$-ray illustrating transitions in band B1.  Inset shows transitions in B1, which are in coincidence with both 1051- and 1474-keV $\gamma$-rays.}\label{b1}
\vspace{0.0cm}
\end{figure}


\section{Experimental details and results}\label{exp.details}

High-spin states of $^{70}$Ge were populated using fusion-evaporation reaction $^{64}$Ni($^{12}$C, $\alpha$2n)$^{70}$Ge.  A beam of $^{12}$C  at 55 MeV energy was delivered by the 15 UD Pelletron accelerator \cite{pel} at Inter University Accelerator Centre, New Delhi.  The target used in this  experiment was an isotopically  enriched $^{64}$Ni with thickness of $\sim$ 1.5 mg/cm$^2$ on a  7 mg/cm$^2$ thick Au backing.  The de-excitation cascades of  $\gamma$-rays from residual nuclei were detected using the Gamma Detector Array (GDA)~\cite{gda}.  The GDA consisted of 12 Compton suppressed n-type Hyper Pure Germanium (HPGe) detectors, having 23\% efficiency relative to 3" x 3" NaI(Tl) crystal. These detectors were arranged in three groups, with each group consisted of four detectors, at angles 50$^\circ$, 98$^\circ$ and 144$^\circ$ with respect to beam direction.  Anti-Compton shields (ACS) made of Bismuth Germanate (BGO) were used to suppress the background from Compton scattered events. 

The data were recorded using an online CAMAC-based data acquisition system called Freedom~\cite{candle1} and trigger was set when at least two detectors were fired in coincidence.  A total of more than 13 $\times$ 10$^7$  twofold or higher coincidence events were recorded in list mode. About 20\% of the recorded events correspond to $\alpha$2n evaporation channel leading to the nucleus of interest $^{70}$Ge.  Offline data analysis was carried out using programs RADWARE \cite{radware}, CANDLE \cite{candle2} and INGASORT \cite{ingasort}. List-mode data were sorted into a E$_\gamma$- E$_\gamma$ matrix from which coincidence spectra were generated with an energy dispersion of 0.5 keV/channel.  In addition, separate 4k x 4k angle-dependent matrices were constructed by taking energies of $\gamma$-ray transitions from all detectors at 50$^\circ$ or 144$^\circ$ on one axis and coincidence $\gamma$ energies from rest of the detectors at 98$^\circ$ on the other axis.  These matrices were used to assign multipolarities of the $\gamma$ transitions using the directional correlation of oriented state (DCO) technique \cite{dco2}. The experimental DCO ratios for the present work is defined \cite{73As} as 

\begin{equation}
R_{DCO} = \dfrac{I_{\gamma_1} \;\;at \;\; 50^{\circ}\;or\;144^{\circ} \;\;gated   \;\;by\;\; \gamma_2 \;\;at\;\; 98^\circ}{I_{\gamma_1} \;\;at\;\; 98^\circ \;\;gated   \;\;by\;\; \gamma_2 \;\;at\;\; 50^\circ\;or\;  144^{\circ}} 
\end{equation}

If the gating transition is of stretched quadrupole nature then this ratio is $\sim$ 1 for pure quadrupole transitions and 0.5 for pure dipole ones.  If the gating transition is of pure dipole multipolarity then this ratio is between 0 to 2 depending on the mixing ratio, and is 1 for pure dipole transitions.

\begin{figure}
\centerline{\includegraphics[scale = 0.37, angle=0 ]{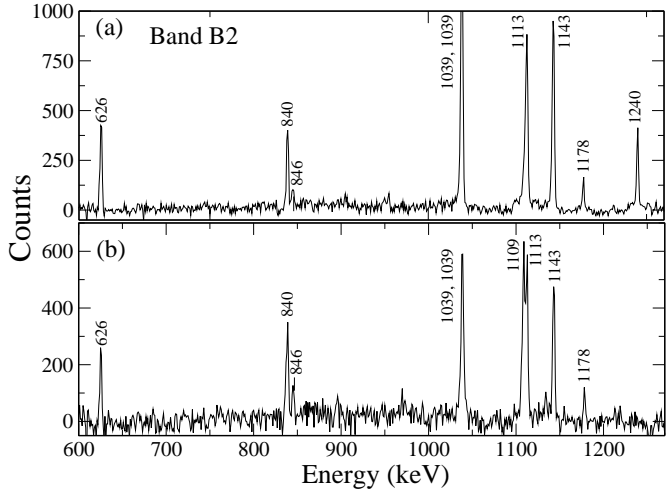}}
\caption{ Representative $\gamma$-$\gamma$ coincidence spectra showing the transitions in band B2, common in gates on (a) both 1134- and 1109-keV, and (b) both 1134- and 1240-keV $\gamma$-rays.} \label{pp2015}
\vspace*{0.1cm}
\end{figure}

The level scheme of $^{70}$Ge has been extended up to the state with J$^\pi$= (20$^+$) and excitation energy 10.2 MeV based on $\gamma$-$\gamma$ coincidence relationships, intensity arguments and DCO ratio measurements.  The partial level scheme of $^{70}$Ge established in the  present work, relevant to the discussion in this article is shown in Fig.\ref{fig:level-scheme}. The ground state positive-parity band determined from the present work is shown in Fig.1 as band B1.   This band was known previously up to spin, J$^\pi$ = (12$^+$) \cite{sugawara,pre2,pre3} and is extended to 14$\hbar$ in the present work with inclusion of 1051 keV $\gamma$-ray transition on top of 12$^+$ level.  An example of $\gamma$-$\gamma$ coincidence spectrum gated on 906 keV is shown in Fig. \ref{b1} illustrating the transitions in band B1. The inset of this figure shows the transitions in band B1 consistently, common in gates on 1051 and 1474 keV $\gamma$-ray transitions.  An important observation in the present work is identification of a new band structure B2  which arises from forking of the ground state band at 6$^+$ state.  Such band structures with forking have also been observed earlier in neighboring nuclei $^{66-68}$Ge \cite{66Ge,68Ge-1, 68Ge-2,68Ge}.  The band B2 is extended to 20$\hbar$ with the addition of five new $\gamma$-transitions of energies 1240, 840, 626, 1178 and 846 keV above the 5538 keV state.  Representative $\gamma-\gamma$ coincidence spectra gated on 1134- and 1109-keV, 1134- and 1240-keV $\gamma$-rays (generated using the AND logic in RADWARE \cite{radware}) are shown in panels (a) and (b) of Fig. \ref{pp2015}, display the newly identified transitions in band B2.  The DCO ratios calculated from two asymmetric matrices for all the transitions in band B2 (except 846 keV, which is quite weak) are consistent with stretched quadrupole nature, and therefore they are placed in the level scheme as $\Delta$J=2 spin sequence.

\begin{figure}
 \vspace{0.4cm}
\includegraphics[width=8.0cm, height=4.5cm, trim = 1cm 0cm 1cm 0cm ]{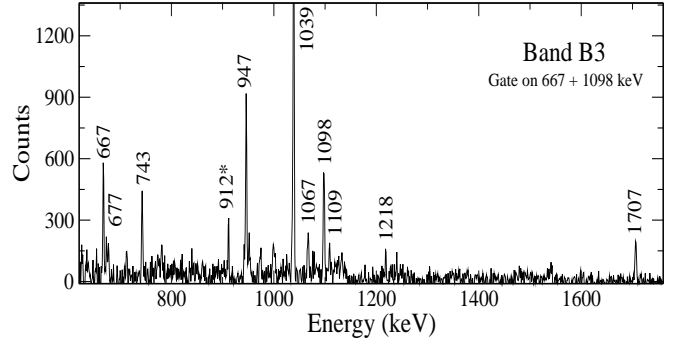}
 \caption{Sum $\gamma$-$\gamma$ coincidence spectrum displaying the transitions in band B3 with gates on 667-, and 1098-keV $\gamma$-rays.  The 912-keV $\gamma$-ray marked with asterick is a contaminant from $^{73}$As.}\label{bandb3}
 \end{figure}

 The band B3 is extended up to spin 8$\hbar$ by placing a 1067-keV $\gamma$ transition above 3752-keV state.  A 1218 keV $\gamma$-transition decaying from (5$^+$) to 3$^+$ state in band B3 is also confirmed in the present work, consistent with the placement in Ref~\cite{pre2}, whereas this transition was not reported in recent work \cite{sugawara}.  A representative sum $\gamma-\gamma$ coincidence spectrum gated on 667- and 1098-keV $\gamma$-rays is shown in Fig. \ref{bandb3}.  Parity for energy levels in bands B1, B2 and B3  are assigned based on earlier works and from systematics \cite{sugawara, 66Ge, 68Ge}. Details of the $\gamma$-ray energies, measured relative intensities, DCO ratios and multipolarities of the observed $\gamma$-ray transitions of $^{70}$Ge are summarized in Table~\ref{DCO}. 
 
\setlength{\tabcolsep}{4pt}
\begin{table}
\centering
\caption{Transition energy (E$_\gamma$), relative intensity (I$_\gamma$), DCO ratio (R$_{DCO}$), multipolarity of the transition (Q: Quadrupole/D: Dipole), and spins of initial (J$_{i}^{\pi}$) and final states (J$_{f}^{\pi}$) for the $\gamma$-transitions shown in the level scheme  of $^{70}$Ge, are listed. Relative intensities are calculated with respect to the 1143-keV transition by normalizing its intensity to a value of 100.  $\Delta$J = 2 transitions are used as gating transitios for DCO ratio measurements.  Errors are given in parentheses for I$_\gamma$ and R$_{DCO}$. Multipolarity mentioned in parenthesis is tentative. \label{DCO}}
\begin{center}

\begin{tabular}{cccccc}
\hline\hline
 \emph{$E_{\gamma}$} & \emph{$I_{\gamma}$} & \emph{$R_{DCO}$}  & \emph{Multipolarity of}  &\emph{$J_i^{\pi}$} &\emph{$J_f^{\pi}$} \\
(keV)   &  (Rel.)       &               & \emph{transition}   &   \\
\hline\hline  
450 & 1.5(3) & - & (Q) & 8$^+$ & 6$^+$\\

490 & 0.8(2) & - & (Q) & 6$^+$ & 4$^+$ \\

626 & 8.9(11) & 1.17(22) & Q & 16$^{+}$ &  14$^{+}$\\

653 & 1.2(4) & - &  - & 4$^+$ & 4$^+$  \\

667 & 11.9(6) & 0.94(7) &$\Delta I=0$, Q & 2$^+$ &  2$^+$\\ 

677 & 1.0(3) & - & (Q) & 8$^+$ &  6$^+$\\ 

743 & 3.4(3) & 0.72(13) & D & 3$^+$ & 2$^+$\\

840 & 10.8(10) & 1.09(18) & Q & 14$^{+}$ &  12$^{+}$\\

846 & 2.2(8) &  & (Q) & (20$^{+}$) &  18$^{+}$\\

906 & 51.1(12) & 0.99(7) & Q & 8$^{+}$ &  6$^{+}$\\

947 & 6.9(5) & 1.01(3) & Q & 6$^+$ &  4$^+$\\

1039 & 183.4(9) & 1.01(5) & Q & 2$^{+}$ & 0$^{+}$\\

1039 & 24.8(9) & 1.12(11) & Q & 10$^{+}$ & 8$^+$\\

1051 & 11.7(14) & 1.17(12) & Q & 14$^{+}$ & 12$^{+}$\\ 

1067 & 3.3(5) & - & (Q) & (8$^+$) &  6$^+$\\

1098 & 9.3(7) & 1.19(10) & Q & 4$^+$ &  2$^+$\\

1109 & 23.1(12) & 0.98(12) & Q & 10$^{+}$ & 8$^{+}$\\

1113 & 134.9(9) & 1.05(6) & Q & 4$^{+}$ & 2$^{+}$ \\

1134 & 29.5(9) & 1.01(9) & Q & 8$^{+}$ & 6$^{+}$ \\

1143 & 100 &  & Q  & 6$^+$ & 4$^+$\\

1178 & 4.3(9) & 0.94(25) & Q & 18$^{+}$ &  16$^{+}$\\

1218 & 1.5(4) & - & (Q) & (5$^+$) &  3$^+$\\

1240 & 14.1(11)  & 1.08(15) & Q & 12$^+$ & 10$^+$\\  

1411 & 3.4(4) & - & (D) & 3$^+$ & 2$^+$\\

1474 & 14.7(11) & 1.13(12) & Q & 12$^+$ & 10$^+$  \\

1707 & 4.7(5) & - & (Q) & 2$^+$ & 0$^+$  \\

\hline\hline

\end{tabular}

\end{center}
\end{table}

\begin{figure}
 \includegraphics[scale=0.43, angle=0]{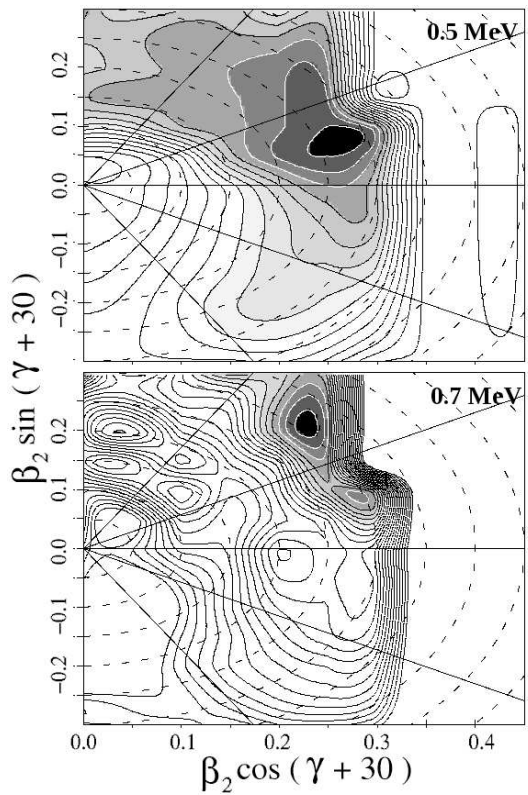}
  \caption{ Total Routhian surface calculations for positive-parity, postive signature states ($\pi $, $\alpha $) = (+, +) \cite{cranking} for $^{70}$Ge  at rotational frequencies of  0.50 MeV (top panel) and 0.70 MeV (bottom panel). The energy separation between adjacent contours is 0.2 MeV.}\label{TRS} 
 \vspace{0.0 cm}
 \end{figure}

\begin{figure}
\includegraphics[scale=0.30,angle=270]{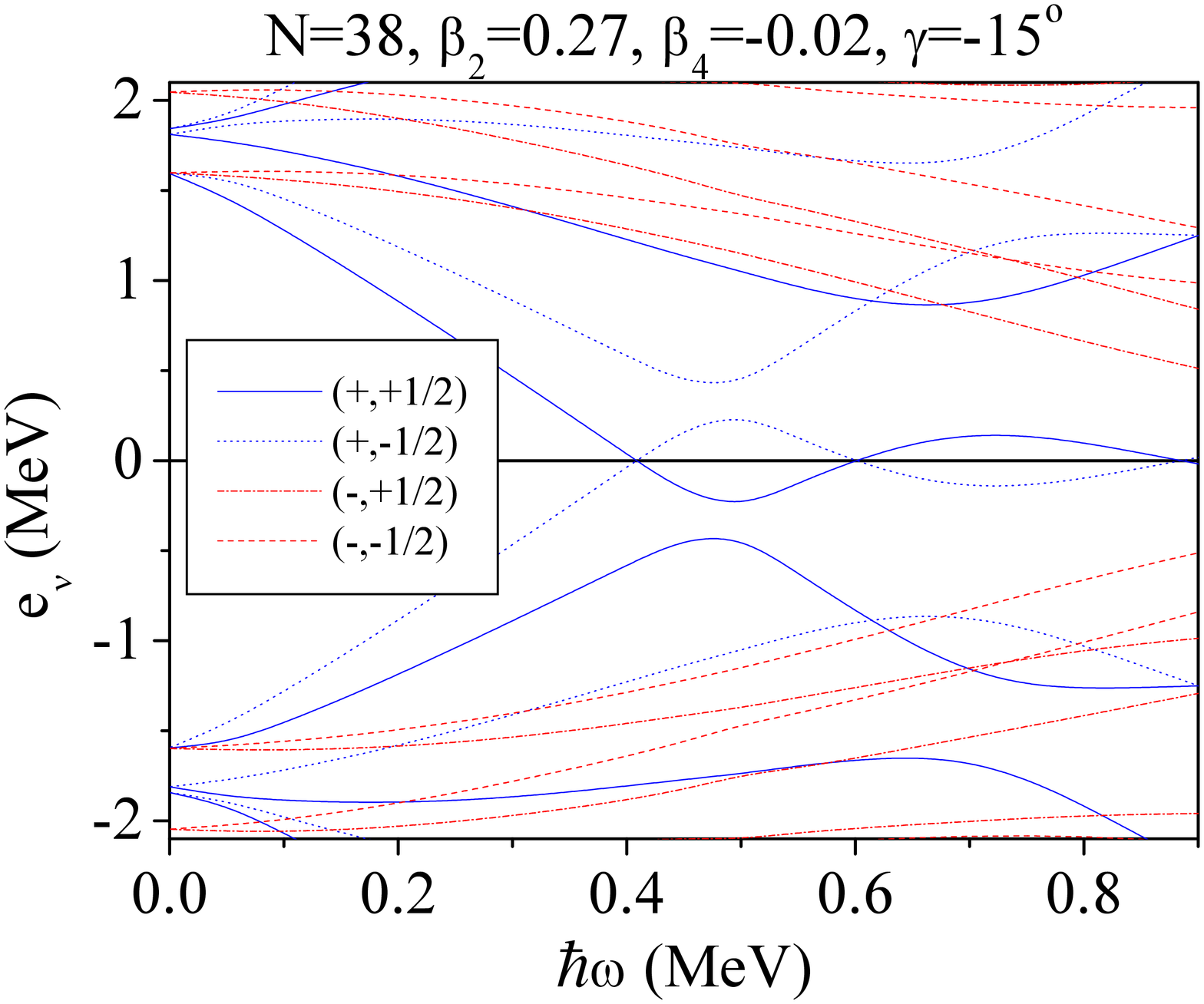}
\vspace{0.2cm}
\includegraphics[scale=0.30,angle=270]{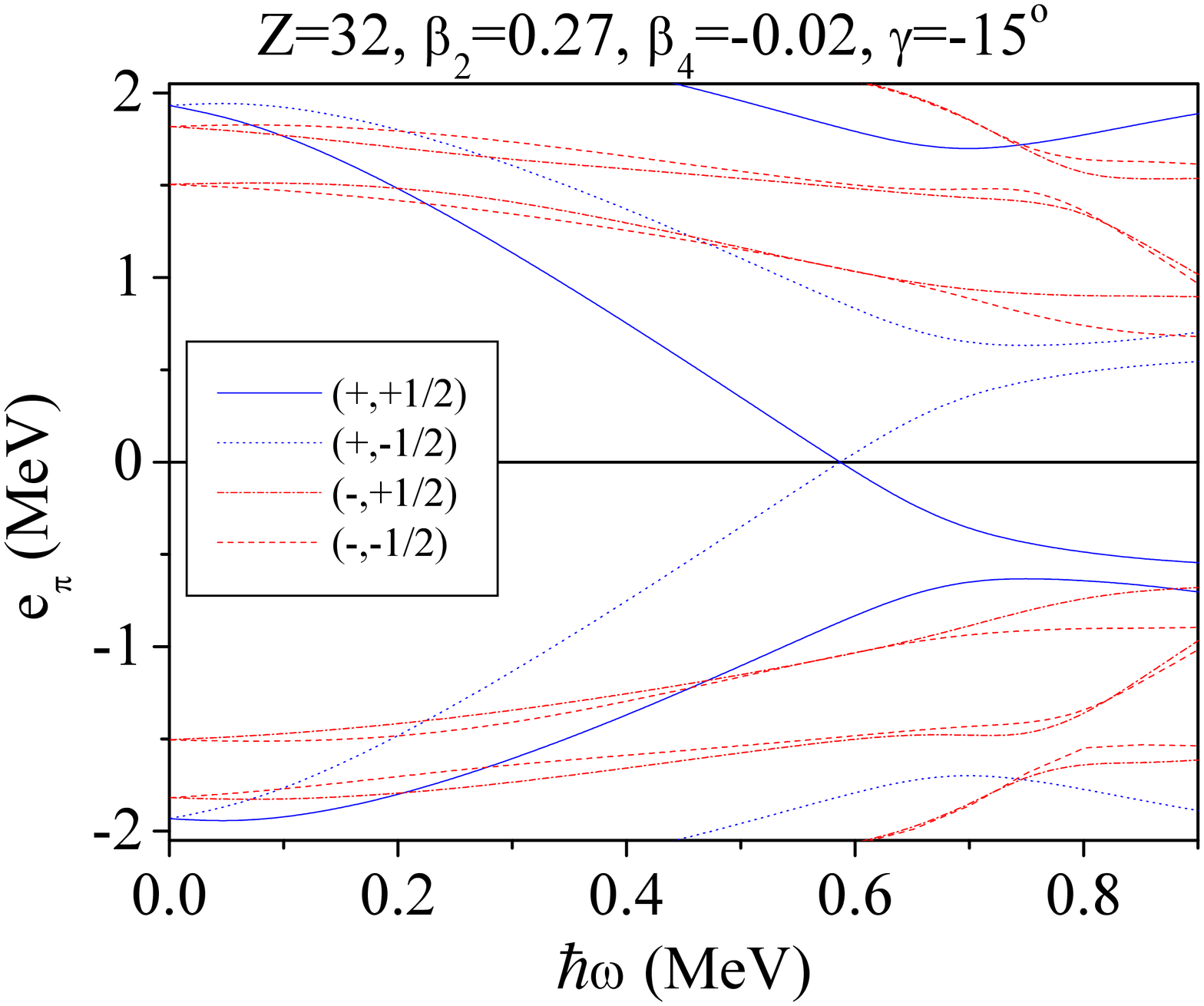}
\caption{\label{qp}(Color online) Cranked shell model calculations using the universal Woods-Saxon potential for quasineutrons (top panel) and quasiprotons (bottom panel) for $^{70}$Ge.}
\end{figure}
 
\section{Discussion}\label{discussion}




Low-spin positive-parity states in $^{70}$Ge were interpreted by several authors using various theoretical models \cite{ibm,ibm2,hfb}. In the present work, the observed band structures and shape evolution are discussed using standard cranked shell model and triaxial projected shell model approaches. 

\subsection{Cranked Hartree Fock Bogoliubov analysis}\label{trs}

\begin{figure}
\includegraphics[scale=0.34,angle=0]{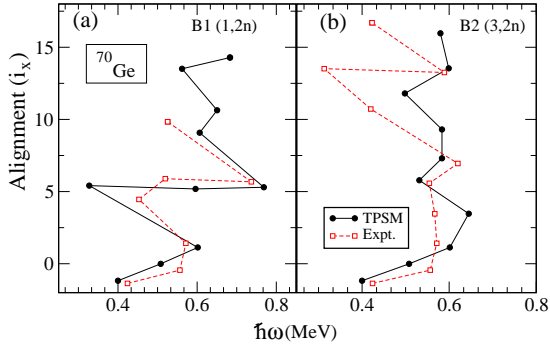}
\caption{\label{fig:alignment}(Color online) Experimental alignments as a function of rotational frequency
for bands, B1 and B2 in $^{70}$Ge.  The reference rotor which was subtracted is based  on Harris parameters, J$_0$ = 6.0 $\hbar^2$/MeV and J$_1$ = 3.5 $\hbar^4$/MeV$^3$. The alignments using the TPSM approach, to be discussed later, are also included for a comparison.}
\vspace{0.3cm}
\end{figure}

  \begin{figure}
 \includegraphics[width=6.0cm, trim = 1cm 0cm 1cm 0cm]{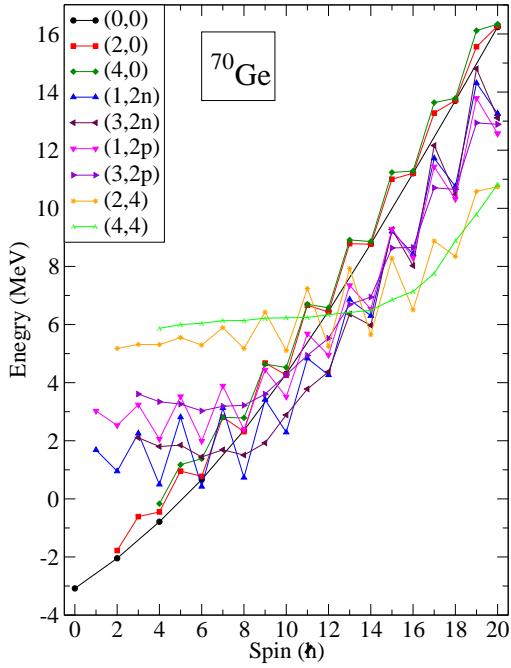}
 \caption{(Color online)Band diagrams for $^{70}$Ge.  Labels  ($K$,n-qp) indicate the $K$-value and the quasiparticle character of  the configuration, for instance, $(3,2n)$ corresponds to the
 $\gamma$-band built on this 2n-aligned state.  For clarity, only the lowest projected K-configurations are shown and in the numerical calculations projections have been performed from
 more than forty intrinsic states. }\label{figtpsm1}
 \vspace{0.0 cm}
 \end{figure}

Hartree-Fock-Bogoliubov cranking calculations have been performed using the universal parametrization of the Woods-Saxon potential with short range monopole pairing~\cite{HFB}.  BCS formalism was used to calculate the pairing gap $\Delta$ for both protons and neutrons. Total Routhian Surface (TRS) calculations were performed in the ($\beta_2$, $\gamma$) plane at different rotational frequencies and the total energy was minimized with respect to hexadecapole deformation ($\beta_4$).  TRS plots for favored positive-parity states (+, +) are shown in Fig.~\ref{TRS} at rotational frequencies of $\hbar\omega$ = 0.5 and 0.7 MeV. These indicate that  the nucleus has substantial quadrupole deformation.  At a rotational frequency $\hbar\omega$ = 0.5 MeV, in the vicinity of the first band crossing, a minimum is seen at ($\beta_2$, $\gamma $) $\approx $ (0.27, -15$^ \circ$), indicating that the nuclear shape is triaxial, but approaching prolate ($\gamma$=0$^\circ$).  At even higher rotational frequency ($\hbar\omega$ = 0.7 MeV), the TRS predict a fairly well-defined minimum with $\gamma \approx$ +12$^\circ$ and approximately the same quadrupole deformation. The energy minimum moves towards increasingly positive values of $\gamma $ at higher rotational frequencies, indicating loss of collectivity.  To investigate the nature of observed bands and crossing frequencies, the quasiparticle routhians were calculated for $\beta_2 \approx$ 0.27, $\gamma \approx$ -15$^\circ$ as a function of rotational frequency \cite{Tandel} and are depicted in Fig.~\ref{qp}. The neutron crossing is predicted at a considerably lower rotational frequency ($\hbar\omega$ = 0.5 MeV), while the proton crossing is expected at a much higher frequency, $\hbar\omega$ = 0.75 MeV.

Cranking formalism \cite{cranking} has been applied to extract the experimental alignments (i$_x$) as a function of rotational frequency ($\hbar \omega$). Figure \ref{fig:alignment} shows the alignment plot for  bands B1 and B2 in $^{70}$Ge.  The observed alignment at $\hbar\omega \approx$ 0.50 MeV for  band B1, shown in Fig.\ref{fig:alignment}, is attributed to $g_{9/2}^2$ neutron alignment, consistent with predictions in previous work ~\cite{sugawara,pre2,pre3}. In comparison to neighboring isotopes,   the observed crossing in band B1 of $^{70}$Ge occurs slightly earlier (by $\approx$ 0.16 MeV) than the observed alignments in $^{66-68}$Ge \cite{66Ge,68Ge}. This might be attributed to the shape change in $^{66-70}$Ge  due to large shell gaps existing at N = 34 and 36 and 38 in the Nilsson single particle level diagram.  The newly identified positive-parity band B2 having band head spin, $I=8\hbar$ also exhibits band crossing around rotational frequency $\approx$ 0.53 MeV (Fig. \ref{fig:alignment}b) with moderate band interaction above the 6$^+$ state which is similar to the one observed in yrast band B1.  The proton band crossing is ruled out as that is expected at $\approx$ 0.75 MeV from the cranked shell model analysis.  Thus the observed band crossings in both the bands, B1 and B2 are attributed to g$_{9/2 }$ neutrons. The second alignment is also  observed in band B2 above spin 14$^+$ which might be composed of four quasiparticle structure.  As is evident from Fig. \ref{fig:alignment} that the observed alignments in both bands B1 and B2 are consistent with the TPSM results which are discussed in the following section.

\subsection{Triaxial projected shell model calculations}

\begin{figure}
 \includegraphics[width=7.8cm, trim = 1cm 0cm 1cm 0cm]{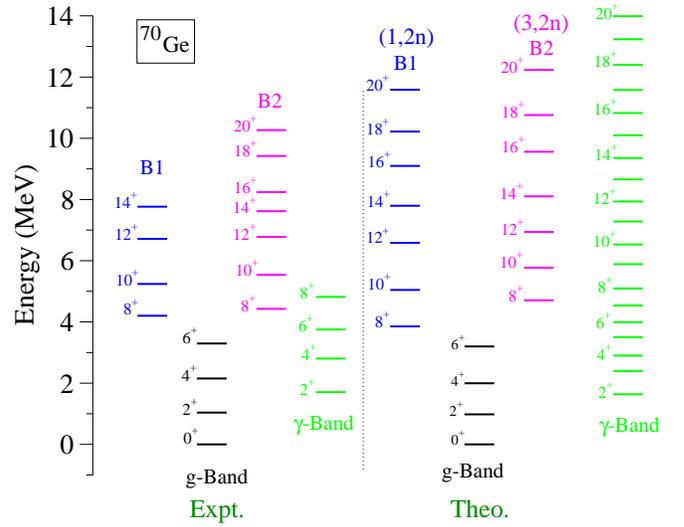}
 \caption{(Color  online) Comparison of calculated  energies by TPSM  with observed experimental data for $^{70}$Ge.}\label{figtpsm2}
 \end{figure}

TPSM Hamiltonian consists of pairing plus
quadrupole-quadrupole interaction terms \cite{KY95} :
\begin{equation}
\hat H = \hat H_0 - {1 \over 2} \chi \sum_\mu \hat Q^\dagger_\mu
\hat Q^{}_\mu - G_M \hat P^\dagger \hat P - G_Q \sum_\mu \hat
P^\dagger_\mu\hat P^{}_\mu,
\label{hamham}
\end{equation}
with the last term in (\ref{hamham}) being the quadrupole-pairing
force. Interaction strengths of the model Hamiltonian are chosen as follows:  $QQ$-force strength $\chi$ is adjusted such that the physical quadrupole deformation $\epsilon$ is obtained as a result of  self-consistent mean-field HFB calculation \cite{KY95}.  Monopole pairing strength $G_M$ is of
the standard form
\begin{equation}
G_{M} = (G_{1}\mp G_{2}\frac{N-Z}{A})\frac{1}{A} \,(\rm{MeV}),
\label{gmpairing}
\end{equation}
 where $- (+)$ is for neutron (proton).

In the present calculation, we use $G_1=20.82$ and $G_2=13.58$,
which approximately reproduces the observed odd-even mass differences
in this region \cite{Chanli15, js01, rp01}.  The oscillator model space considered in the present
work is $N=3, 4$ and 5 for both neutrons and protons.  The quadrupole pairing strength $G_Q$ is
assumed to be proportional to $G_M$ and the proportionality constant fixed to 0.18.  These interaction strengths are consistent with those used earlier for the same mass region \cite{JS14}.  Intrinsic quasiparticle states have been constructed for $^{70}$Ge with the deformation parameters of $\epsilon = 0.235$ and $\epsilon'=0.145$ \cite{JS14}.

The projected states from various intrinsic states close to the Fermi surface are displayed in Fig.~\ref{figtpsm1}.  The ground-state, $\gamma$- and $\gamma\gamma$-bands labeled by $(0,0), (2,0)$ and $(4,0)$ result  from angular-momentum projection of the vacuum configuration by specifying K=0, 2 and 4 respectively in the rotational D-matrix \cite{RS80}.  It is noted that $\gamma$- and $\gamma\gamma$-bands depict a substantial signature splitting and even-spin states of the $\gamma$-band are close in energy to the  ground-state band. 

\begin{figure}
\includegraphics[width=6.5cm, trim = 1cm 0cm 1cm 0cm]{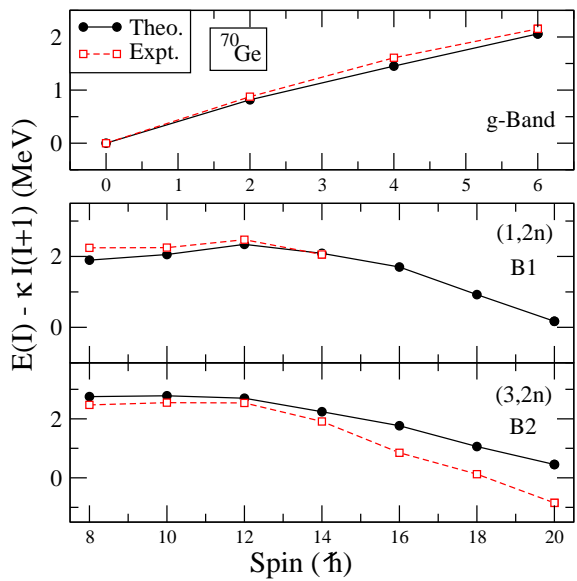}
\caption{(Color online) Comparison of calculated  energies with observed experimental level energies subtracted from reference value displayed as a function of spin for the band B1, B2 and ground-state band (g-band) in $^{70}$Ge.}\label{EvsI-70ge}
\end{figure}

What is most interesting to observe from Fig.~\ref{figtpsm1} is the crossing of K =1 two neutron-aligned configuration $(1,2n)$ with the ground-state band at spin, I=6$\hbar$. Further the $\gamma$-band built on this configuration with K =3 also crosses the ground-state band between I=6 and 8$\hbar$. These aligning states result from the projection of the same intrinsic state but having different K-values.  Since K=3 two-neutron aligned $\gamma$-band has lower signature splitting as compared to the parent band, the lowest odd-spin members along the yrast-band shall originate from this configuration.  The proton-aligned configurations, $(1,2p)$ and $(3,2p)$ lie at higher excitation energies and do not cross the ground-state band.  However, the two-neutron plus two-proton aligned configurations crosses the two-neutron aligned configuration above I=14$\hbar$ and the yrast-band above this spin value is composed of four-quasiparticle states.


\begin{figure}
\includegraphics[width=6.5cm, trim = 1cm 0cm 1cm 0cm]{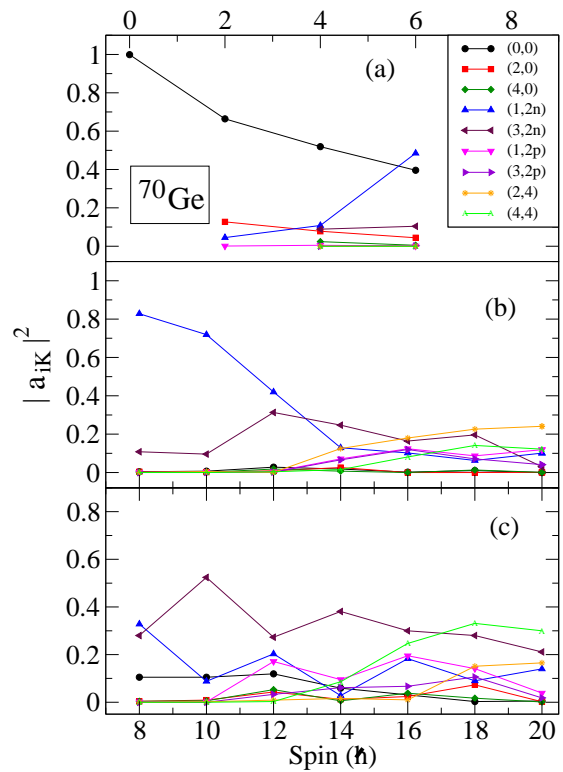}
\caption{(Color online) Probability of various projected K-configurations in the  wavefunctions of the observed  bands for $^{70}$Ge. See caption of  Fig. \ref{figtpsm1} for meaning of various symbols.}\label{figtpsm3}
\vspace{0.2 cm}
\end{figure}


Projected states shown in Fig.~\ref{figtpsm1} and many more states around the Fermi surface ($\sim 40$ in number ) are then employed to diagonalize the shell model Hamiltonian, Eq.~(\ref{hamham}).  Energies obtained after  diagonalization are compared with the experimental data in Fig.~\ref{figtpsm2}.  It is evident from the figure that the experiential data is reproduced reasonably well by TPSM calculations.  This can seen more clearly in Fig.~\ref{EvsI-70ge} where experimental data are compared with TPSM calculations for ground-state, B1 and B2 bands after subtracting the level energies from reference value.  The experimental level energies degenerate with TPSM results up to highest observed spin in band B1. In case of band B2, level energies are almost degenerate up to spin 14$\hbar$ and then deviate at higher spin. This could be due to shape changes at higher spins as predicted by the TRS study presented in Sec.~\ref{trs}.

Further, to probe the structure of  high-spin states shown in Fig.~\ref{figtpsm2} after band mixing, dominant components of projected wavefunctions of the states are displayed in Fig.~\ref{figtpsm3}.  The ground-state band up to spin, I= 4$\hbar$ has predominately 0-quasiparticle configuration with K=0 as is evident from the panel (a) of Fig.~\ref{figtpsm3}.  The spin state with I=6$\hbar$ has substantial contribution from the two-quasiparticle neutron configuration having K=1.  The amplitudes of the wavefunctions of two aligned bands observed above ground-state band are shown in panels (b) and (c). These are noted to be dominated by K=1 two-neutron aligned configuration $(1,2n)$ and the $\gamma$-band built on this aligned state with K=3 for angular-momentum states of I=8, 10, 12 and 14$\hbar$.  For high-spin states of I=16$\hbar$ and above, the wavefunctions are mostly composed of four-quasiparticle states.

\subsection{Comparison with band structures in $^{68}$Ge}

The nature of observed quasiparticle alignments and band structures in $^{70}$Ge is quite similar to its neighboring isotope $^{68}$Ge~\cite{68Ge-1,68Ge-2,68Ge}, in which the ground state band forks beyond $I=8^+$ into  multiple band structures.  

\begin{figure}
\includegraphics[width=7.5cm, trim = 1cm 0cm 1cm 0cm]{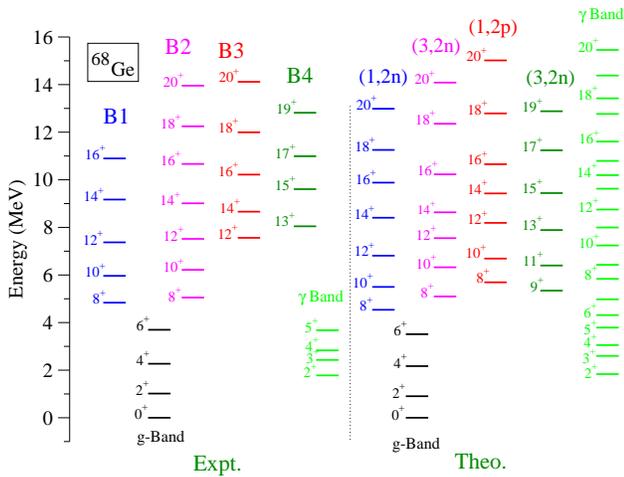}
\caption{(Color online) Comparison of the calculated TPSM  energies with available experimental data for $^{68}$Ge \cite{68Ge}.}\label{figtpsm4}
\end{figure}

\begin{figure}
\includegraphics[scale=0.38,angle=0]{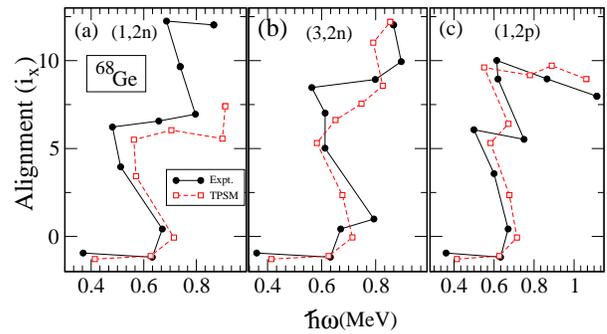}
\caption{\label{fig:alignment68ge}(Color online) Comparison of experimental and TPSM alignments as a function of rotational frequency for the bands B1, B2 and B3 in $^{68}$Ge.}
\vspace{0.3 cm}
\end{figure}

To have an insight of the nature of observed alignments and forking band structures in $^{68}$Ge, we have  performed the TPSM calculations for $^{68}$Ge with deformation parameters of $\epsilon = 0.22$ and $\epsilon'=0.16$.  The predicted TPSM band structures after band mixing are compared with experimental data  in Fig.~\ref{figtpsm4}. It is evident from the figure that the four observed bands, B1 to B4 above ground state band are reproduced quite well by  TPSM approach.  Figure \ref{fig:alignment68ge} shows the comparison of observed alignments with TPSM calculations as a function of rotational frequency for the bands B1, B2 and B3, indicating that all three bands have sharp band crossings and are composed of two-quasiparticle structures after the bandcrossing.  Further, from the analysis of the TPSM wavefunctions, it is seen that these three bands B1, B2 and B3 have  dominant structure of two-neutron aligned band with K=1, $\gamma$-band built on this aligned  configuration having K=3 and two-proton aligned band with K=1  respectively.  The odd-spin band has dominant contribution from  $\gamma$-band built on the neutron-aligned band with K=3.  Therefore, the two even-spin aligned band structures are predicted to have same neutron configuration and the third one has proton structure.  In previous work, these three even-spin bands B1, B2 and B3 have been interpreted \cite{68Ge-2, 68Ge} as  two two-neutron aligned bands and the configuration of the third band  remained unresolved. The g-factor measurements of the states in $^{70}$Ge and $^{68}$Ge are highly desirable to probe further the predicted intrinsic structures of observed bands.

\section{Summary and conclusions}\label{summary}

In summary, a new positive-parity band has been identified in $^{70}$Ge through $\gamma$-ray spectroscopic study which extended the level scheme up to (20$\hbar$) and excitation energy of 10.2 MeV.  The intensity of the ground state band forks into two branches  above  6$^+$ state, resulting into two positive-parity band structures.  It has been demonstrated using CSM and TPSM approaches that both the observed band structures have two-neutron aligned configurations.  The possibility of proton structure is ruled out as in CSM study it occurs at $\hbar \omega=0.75$ MeV and in both the bands the crossing is observed at $\hbar \omega \approx 0.5$ MeV. From the TPSM wavefunctions,  it is noted that band B1 is based on two-neutron quasiparticle configuration having K=1 and band B2 is predicted to be a $\gamma$-band built on this aligned two-quasiparticle band with K=3.  The forking of the ground-state band into two bands in $^{70}$Ge has, therefore, a different origin as compared to the earlier  observed forking in nuclei.  Further, it has been shown that one of the observed bands in $^{68}$Ge also has the structure of the $\gamma$-band built on the two-quasipaticle configuration, indicating that this kind of two-quasiparticle band structures may be more widespread and need to be explored in other nuclei and mass regions of the periodic table. 

\section{Acknowledgments}
We thank the Pelletron crew of the IUAC, New Delhi for their support during the experiment and the target laboratory group of IUAC for their help in making $^{64}$Ni target. We also thank Professor S. C. Pancholi for his valuable suggestions.  The author (M.K.R) would like to acknowledge financial support provided by UFR project fellowship (No. 42328), IUAC, New Delhi and Senior Research  Fellowship (No.09/002(0494)/2011-EMR-1) from Council of Scientific and Industrial Research (CSIR), India.

\end{document}